\documentclass[aps,prc,preprint,tightenlines,superscriptaddress,showpacs,%
amssymb,byrevtex,nofootinbib]{revtex4}
\usepackage[dvips]{graphicx,color}
\usepackage{amsmath}
\usepackage{mathrsfs}
\usepackage{dcolumn}
\usepackage{epsfig}
\usepackage{bm}

\begin{document}
%\title{Spectra from Realistic $\pi N$ and  $N N$ Hamiltonians for testing LQCD}
\title{From Extraction of Nucleon Resonances to LQCD } 
%%%%%%%%%%%%%%%%%%%% Authors %%%%%%%%%%%%%%%%%%%%%%%%%%%%%%%%%
\author{T.-S. H. Lee}
\affiliation{Physics Division, Argonne National Laboratory, Argonne, Illinois 60439, USA}
\author{Jia-Jun Wu}
\affiliation{Special Research Center for the Subatomic Structure of Matter (CSSM), School of Chemistry and Physics, University of Adelaide Adelaide 5005, Australia}
\author{Hiroyuki Kamano}
 \affiliation{Research Center for Nuclear Physics, Osaka University, Ibaraki,
Osaka 567-0047, Japan}

\begin{abstract}
The intrinsic difficulties in extracting the hadron resonances  from reaction data                         
are illustrated by using  several exactly soluble $\pi\pi$ scattering models.
The finite-volume Hamiltonian method is applied to predict
spectra using two meson-exchange Hamiltonians of $\pi N$ reactions.
Within a three-channel 
model with $\pi N$, $\pi\Delta$ and $\sigma N$ channels, we show the advantage
of the finite-volume Hamiltonian method over the approach using the L\"uscher formula
to test Lattice QCD calculations aimed at predicting nucleon resonances.
We discuss the necessary steps for using the ANL-Osaka eight-channel Hamiltonian
to predict the spectra for testing the LQCD calculations for  
determining the excited nucleon states up to invariant mass $W= 2 $ GeV.

\end{abstract}
\maketitle

\section{Introduction}
\label{intro}
The excited nucleons are unstable and coupled with the meson-nucleon continuum to form nucleon resonances
($N^*$). 
Thus the properties of the excited nucleons can only be studied by analyzing  the nucleon
resonances extracted from the data of meson production reactions induced by pions, photons and electrons.
The extraction of nucleon resonances has a long history. In recent years, the main advance is to
develop multi-channel approaches to extract nucleon resonances by fitting $simultaneously$
the  data of
$ \pi N, \gamma N \rightarrow \pi N, \eta N, KN,\pi\pi N$, where
the $\pi\pi N$ states contain resonance components $\pi\Delta, \rho N, \sigma N$.
It has been observed in Ref.\cite{anl-osaka}
that three such multi-channel analyses\cite{anl-osaka,bg,juelich} agree well for
the $N^*$ states with energies below about 1.6 GeV, but disagree very significantly at higher energies.
%This is illustrated in Fig.\ref{fg:spectrum} for the states associated with the $P_{33}$ and $P_{11}$ 
%partial waves of pion-nucleon scattering.
In the first part of this contribution, we use several exactly soluble $\pi\pi$ scattering
models to investigate the
sources of these differences and discuss how a dynamical approach used in
the ANL-Osaka\cite{anl-osaka} analysis 
can help reduce the uncertainties in extracting the nucleon resonances.

In addition to extracting the nucleon resonances, the main outcome of the ANL-Osaka analysis is 
a multi-channel model Hamiltonian which can be used to interpret the extracted resonance parameters and
to make predictions for future experimental tests.
In this contribution, we further demonstrate that
the ANL-Osaka multi-channel Hamiltonian can be
used  to relate the spectrum from the LQCD calculations to the nucleon resonances embedded in the 
experimental data of $\pi N$ reactions. This is achieved by applying
the finite-volume Hamiltonian method developed in Refs.\cite{ad-1,ad-2}.

The results from examining the model dependence of the  resonance extraction
 are presented in section 2. The applications
of the finite-volume Hamiltonian method to predict spectra from the ANL-Osaka model
 Hamiltonians will be presented in section 3.
A summary and the discussions on future directions are  given in section 4.
%\begin{figure}[h]
%\begin{center}
%\includegraphics[width=0.5\columnwidth]{spectrum.eps}
%\caption{Spectrum of nucleon resonances in $P_{11}$ and $P_{33}$ states of $\pi N$ scattering}
%\label{fg:spectrum}
%\end{center}
%\end{figure}

\section{Resonance Extraction}
It has been well established\cite{ dalitz,bohm} that
resonances are the eigenstates of the Hamiltonian of the underlying fundamental theory with 
outgoing boundary condition and are associated with the poles of the scattering amplitudes in the 
complex energy($E$)-plane. The extraction of resonances consists of two steps:
\begin{enumerate}
\item Determine the partial-wave amplitudes (PWA)  from the available data.
\item Analytically continue the determined PWA to the complex-E plane for extracting
      the poles and residues that are  near the physical region.
\end{enumerate}
For the step 1, it has been shown\cite{tabakin}
 that the partial-wave amplitudes can be determined up to an overall phase
from the independent  observables obtained by performing
$complete$ experiments.  For example,
for determining the PWA of the 
single pseudo-scalar meson
photo-production we need\cite{shkl} to have  data for the 
differential cross sections ($d\sigma/d\Omega$), single polarizations
($T, P,\Sigma$), and double polarizations
($O_{x'}, O_{z'})$ with linearly polarized photons and
$(C_{x'}, C_{z'})$ with circularly polarized photons.
The measurements should cover all angles at each energy and have high 
accuracy.

In reality, complete experiments are not available and hence
the data for determining PWA are always incomplete and have large statistical 
and systematic errors in some angles or energies. Even the data are
complete, many solutions in the determinations of PWA are possible, mainly
due to the intrinsic difficulty that the cross sections are related to the
amplitudes bi-linearly; i.e.
 $d\sigma/d\Omega = |f^R(\theta)+if^I(\theta)|^2$. This was demonstrated
in a study of the data of $\gamma p \rightarrow K^+ \Lambda$ from
Jefferson Laboratory (JLAB). The details have been presented in Ref.\cite{shkl},
and will not be covered here. Similar  difficulties are also encountered in the
determinations of the partial-wave amplitudes of $\pi\pi$, and $\pi N$ scattering which
will be discussed in this contribution. 

Once the PWA are determined, we then take step 2 to analytically continue the PWA to the
complex-E  plane. This can only be done within a model and thus the model dependence of
resonance extraction is an important issue.
In this section, we use several exactly soluble $\pi\pi$ scattering models
to illustrate this intrinsic difficulty of the resonance extraction.

Following the coupled-channel Hamiltonian formulation of Ref.~\cite{msl}, we assume that
 $\pi\pi$ scattering can be described by vertex interactions $g_{i,\alpha}$, 
which define the decay
of the $i$-th bare state $\sigma_i$ into a two-particle state $\alpha$, and
two-body potentials $v_{\alpha,\beta}$, where $\alpha,\beta =\pi\pi, K\bar{K}$.
In each partial wave, the scattering amplitude is then defined
by the following coupled-channel equations
\begin{eqnarray}
T_{\alpha,\beta}(k,k'; E) &=& V_{\alpha,\beta}(k,k')
+ \sum_{\gamma}\int _0^{\infty} k^{''\,\,2}dk^{''}
V_{\alpha,\gamma}(k,k'')G_{\gamma}(k^{''};E)T_{\gamma,\beta}(k^{''},k';E)\,,
%V_{\alpha,\gamma}(k,k'')\frac{1}{E-E_{\gamma_1}(k^{''}) - E_{\gamma_2}(k^{''})
%+i\epsilon} T_{\gamma,\beta}(k^{''},k';E)\,,
\label{eq:lseq-1}
\end{eqnarray}
where $G_\gamma(k;E)=1/(E-E_{\gamma_1}(k) - E_{\gamma_2}(k) +i\epsilon)$, and 
\begin{eqnarray}
V_{\alpha,\beta}(k,k') = \sum_{i=1,n}g^*_{i,\alpha}(k)\frac{1}{E-m_i^0} g_{i,\alpha}(k')
+v_{\alpha,\beta}(k,k')\,.
\label{eq:lseq-2}
\end{eqnarray}
Here $m_i^0$ is the mass of the $i$-th bare particle.

To proceed, we need to choose the forms of the interactions in
Eq.(\ref{eq:lseq-2}).
The  $\sigma_i \rightarrow\pi\pi,  K\bar{K}$ vertex functions are defined as:
\begin{eqnarray}
g_{i, \alpha}(k)&=&\frac{g_{i,\alpha}}{\sqrt{m_\pi}}
f(c^i_\alpha, k),\label{eq:g}
\end{eqnarray}
where $m_\pi$ is the mass of $\pi$.
The two-particle potentials are assumed to take
   the following
separable form
\begin{eqnarray}
v_{\alpha,\beta}(k, k')
&=&\frac{1}{m^2_\pi}\sum_{m,n}h_{\alpha,m}(d_{\alpha,m}, k)\,\,G^{m,n}_{\alpha\beta}
\,\,h_{\beta,n}(d_{\beta,m}, k'),
\label{eq:sepa-v}
\end{eqnarray}
We will use various models to extract the resonance parameters.
Their differences are in the form of the form factors, the number of
the bare states $\sigma$,  and the number of terms in the separable potentials.
For form factors, we consider the following three parametrizations:
\begin{eqnarray}
A\,\,\,\,:\,\,\,\, f(c^i_\alpha, k)&=&\frac{1}{(1+(c^i_\alpha  k)^2)}\,\,\,\,;\,\,\,
h_{\alpha,m}(d_{\alpha,m}, k)=\frac{1}{(1+(d_{\alpha,m} k)^2)^2}.
\label{eq:parm-a}\\
B\,\,\,\,:\,\,\,\,f(c^i_\alpha, k)&=&\frac{1}{(1+(c^i_\alpha  k)^2)^2}\,\,\,\,;\,\,\,
h_{\alpha,m}(d_{\alpha,m}, k)=\frac{1}{(1+(d_{\alpha,m} k)^2)^4}. \label{eq:parm-b} \\
C\,\,\,\,:\,\,\,\,f(c^i_\alpha, k)&=&e^{-(c^i_\alpha  k)^2}\,\,\,\,;\,\,\,
h_{\alpha,m}(d_{\alpha,m}, k)=e^{-(d_{\alpha,m} k)^2}.\label{eq:parm-c}
\end{eqnarray}
We  consider three models:
\begin{enumerate}
\item Model I-A : The parametrization A is used and
 $m=n=1$ is set to define the potential Eq.(\ref{eq:sepa-v}).
%It has 10 parameters.

\item Model I-B : The parametrization B is used and
   $m,n=1,2$ is set to define the potential Eq.(\ref{eq:sepa-v}).
%This model has 19 parameters.

\item Model I-C : The parametrization C is used and  $m,n=1,2$ is set 
to define the potential Eq.(\ref{eq:sepa-v}).
%This model has 19 parameters.
\end{enumerate}

\begin{figure}[htbp] \vspace{-0.cm}
\begin{center}
\includegraphics[width=0.5\columnwidth]{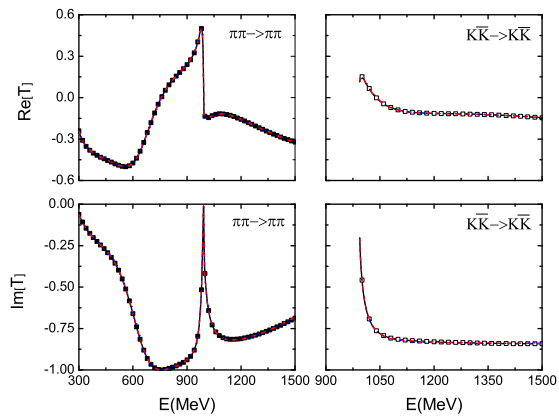}
\caption{The fits to the  $\pi\pi$ and 
$K\bar{K}$ amplitudes generated from Model I-A(solid black)
by using Model I-B (red dashed) and Model I-C(blue dotted). 
They agree within $1\%$ and hence are not distinguishable. }
\label{fg:IBC}
\end{center}
\end{figure}

We first adjust the parameters of Model I-A to roughly fit the data of $\pi\pi$ amplitudes up to about
1 GeV. This model is then used to generate the amplitudes as 'data' in the fits by using the other
two models. We assign extremely small error 1 $\%$  in the fits. Thus the resulting
three amplitudes are almost indistinguishable, as seen in Fig.\ref{fg:IBC}.
Their pole positions and residues are compared in Table \ref{tab:IABC}. 
We see that three models agree extremely well except the residue for $K\bar{K}$ channel
 of the
first resonance near $640$ MeV which is well below the threshold $900$ MeV of the
$K\bar{K}$ threshold.  Apart from this, we conclude that the extraction of resonance
parameters are
independent of the form of the form factors as far as the data are fitted $exactly$.
The small differences seen in Table \ref{tab:IABC} are due to the remaining small discrepancies
between the three models in their resulting amplitudes.
Our results also suggest that the residues of a given channel for the resonance poles which
are well below
the threshold of that channel are not meaningful.

\begin{table}[ht]
     \setlength{\tabcolsep}{0.15cm}
\begin{center}\caption{The pole positions(MeV) and residue(MeV$^{-1}$) of Models I-A, I-B, I-C.}
\begin{tabular}{cccc}\hline
  Model                & Pole Position       &  Residue of $\pi\pi$          & Residue of $K\bar{K}$ \\
 II sheet-1            &                     &    $\times10^{-4}$            &   $\times10^{-4}$     \\
 I-A (data)            & $639.3-i158.9$      &  $5.295-i2.153$               & $12.63+i8.477$       \\
 I-B                   & $637.8-i159.9$      &  $5.368-i2.285$               & $-9.332+i2.054$      \\
 I-C                   & $634.5-i156.2$      &  $5.076-i2.556$               & $191800-i65380$       \\
\hline
&&& \\
 II sheet-2            &                     &    $\times10^{-5}$            &   $\times10^{-5}$     \\
 I-A (data)            & $1000.30-i8.89$     &  $-3.514-i3.088$              & $1.822+i33.81$       \\
 I-B                   & $1000.14-i8.88$     &  $-3.493-i3.111$              & $2.140+i34.62$       \\
 I-C                   & $1000.04-i8.83$     &  $-3.467-i3.162$              & $2.955+i35.39$       \\
\hline\end{tabular}  \label{tab:IABC}
\end{center}
\end{table}

\begin{figure}[htbp] \vspace{-0.cm}
\begin{center}
\includegraphics[width=0.3\columnwidth]{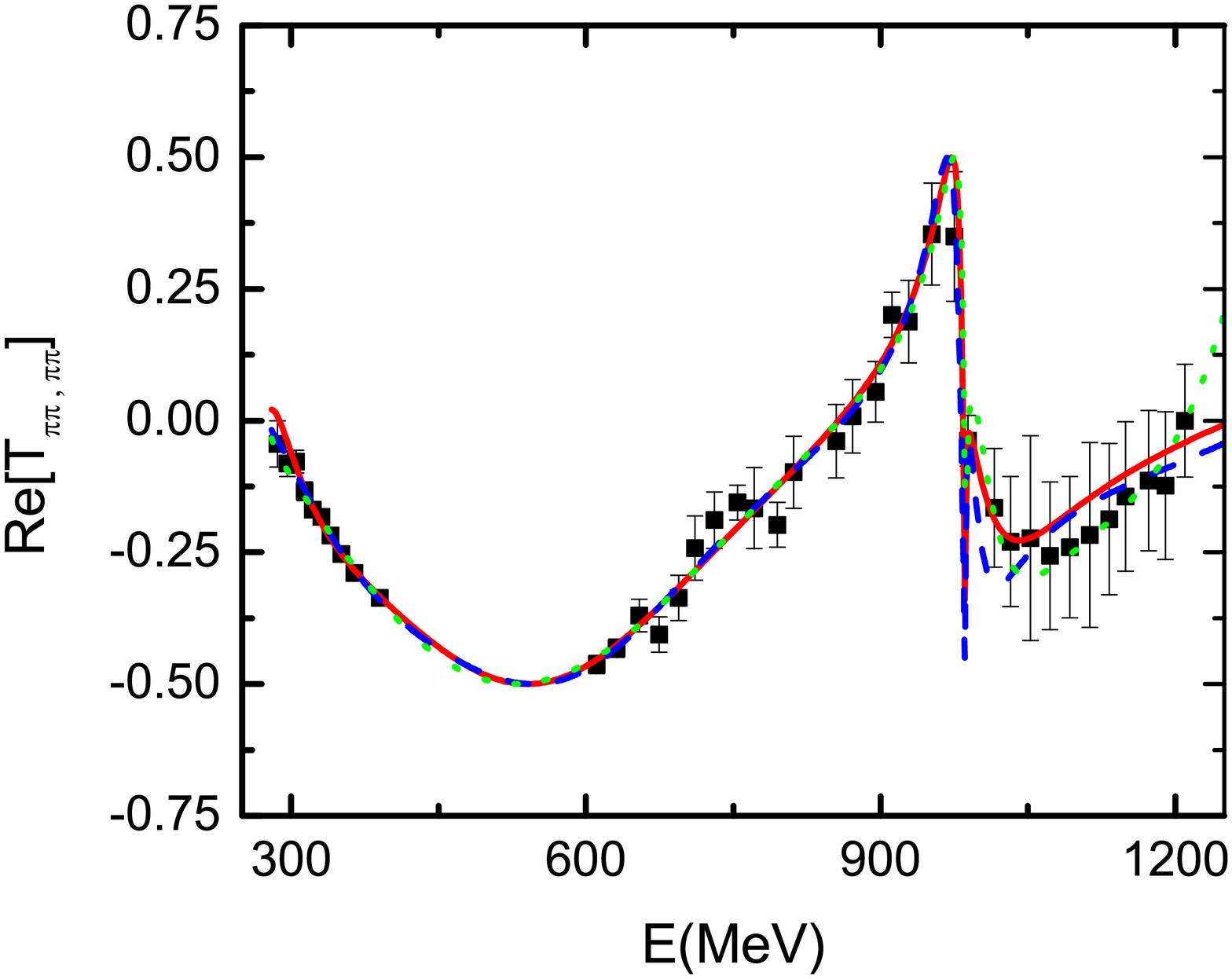}
\includegraphics[width=0.3\columnwidth]{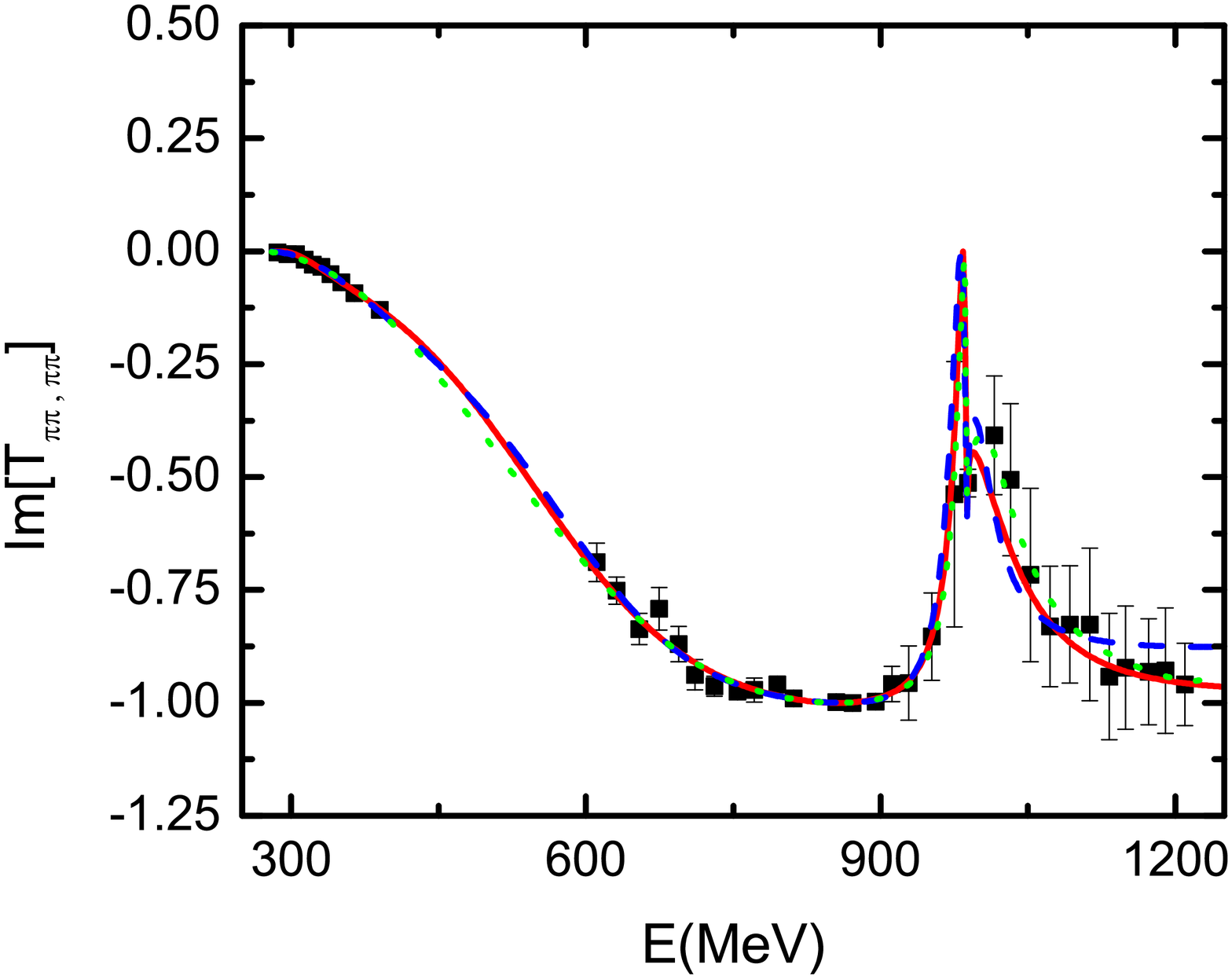}
\caption{The fits to the empirical $\pi\pi$ amplitude 
up to 1.2 GeV. The solid red, dashed blue and dotted magenta lines are 
from using  Models II-A, II-B and II-C.}
\label{fg:exp}
\end{center}
\end{figure}

In reality, the available $\pi\pi$ data have errors and incomplete.
We now examine the extent to which the current $\pi\pi$ data can
determine the resonance parameters. Here we consider three
models with two bare $\sigma$ states and
an one-term separable potential defined by setting
$n=m=1$ in Eq.(\ref{eq:sepa-v}). Their differences are from
using the three different parametrizations specified 
in the Eqs.(\ref{eq:parm-a})-(\ref{eq:parm-c}). They are denoted as
model II-A(Eq.(\ref{eq:parm-a})), II-B(Eq.(\ref{eq:parm-b})), 
and II-C(Eq.(\ref{eq:parm-c}).
We focus on the first two resonances and hence only need to
fit the data up to 1.2 GeV. 
The results are shown in the Fig.\ref{fg:exp}.
Clearly all three models can fit the data equally well within the errors of the data.
The extracted pole positions and residues are listed in Table \ref{tab:IIABC}.
Here we see that the results extracted from three models do not agree well; 
in particular the residues of $K\bar{K}$. This is not surprising since there are no
data for $K\bar{K} \rightarrow K\bar{K}$ amplitudes to constrain the fits.
\begin{table}[ht]
     \setlength{\tabcolsep}{0.15cm}
\begin{center}\caption{The pole positions(MeV) and residue(MeV$^{-1}$) of Models II-A, II-B, II-C.}
\begin{tabular}{ccccc}\hline
  Model         &$\chi^2$  & Pole Position       &  Residue of $\pi\pi$   & Residue of $K\bar{K}$ \\
 II sheet-1     &          &                     &    $\times10^{-4}$     &                        \\
 II-A           &$40$      & $523.7-i264.6$      &  $10.78-i9.323$        & $1.183-i2.595(\times10^{-2})$      \\
 II-B           &$36$      & $597.0-i217.1$      &  $6.157-i3.573$        & $3.198+i3.272(\times10^{-3})$      \\
 II-C           &$43$      & $672.3-i292.0$      &  $5.753+i2.102$        & $2.198+i8.268(\times10^{23})$       \\
\hline
&&&& \\
 II sheet-2     &          &                     &    $\times10^{-5}$     &   $\times10^{-4}$     \\
 II-A           &          & $992.7-i9.73$       &  $-6.356-i3.709$       & $-10.83+i0.3889$       \\
 II-B           &          & $986.6-i15.25$      &  $-6.284-i1.020$       & $4.588+i7.788$         \\
 II-C           &          & $998.5-i11.21$      &  $-8.870-i0.9770$      & $-15.51+i2.208$
 \\\hline\end{tabular}  \label{tab:IIABC}
\end{center}
\end{table}

\section{ANL-Osaka dynamical coupled-channel model and LQCD}
The results discussed in section II indicate that 
in reality the determination of PWA and resonance
extraction cannot be performed model independently.
It is desirable to investigate nucleon resonances  within a reaction model 
that is constrained by the 
well-established physics. Thus  the meson-exchange mechanisms,
which have been well established in the studies of $NN$ scattering\cite{mach}
and $\pi N$ and $\gamma N$ reactions in the $\Delta$ (1232) region\cite{sl},
are used to define the analytic
structure of the
non-resonant reaction amplitudes in the ANL-Osaka analysis.
The results of the ANL-Osaka analysis have been given in detail in Ref.\cite{anl-osaka} and
will not be covered here. Instead, we now turn to explaining how the
finite-volume Hamiltonian method developed in Refs.\cite{ad-1,ad-2} can be applied to
relate the ANL-Osaka model
Hamiltonian to the LQCD calculations.

In a periodic volume
characterized by side length $L$,
the quantized three momenta of mesons and baryons
must be $k_n = \sqrt{n}\frac{2\pi}{L}$ for
integers $n = 0,1,2,\ldots$.
For a given choice of  momenta $(k_0, k_1,\ldots, k_{N-1})$,
solving the Schrodinger equation $(H_0+H_I)|\Psi_E\rangle = E|\Psi_E\rangle $
in finite volume 
is equivalent to finding the eigenvalues  of the
following matrix equations
\begin{eqnarray}
\det ([H_0]_{ N_c+1} + [H_I]_{N_c+1} - E[I]_{ N_c+1}) = 0\,,
\label{eq:det-1}
\end{eqnarray}
where $N_c=n_c\times N$ with $n_c$ denoting the number of channels,
$[I]_{N_c+1}$ is
an $(N_c +1)\times(N_c+1)$ unit matrix.
The essence of the finite-volume Hamiltonian method is  that
the scattering amplitudes
calculated from the predicted  spectrum 
by using the L\"uscher formula\cite{luch-1,luch-2,luch-3} 
are identical to
the scattering amplitudes calculated directly from the Hamiltonian
in infinite volume.
It follows that the spectrum from solving Eq.(\ref{eq:det-1}) at any $L$ 
can be
used as the "data"  to  test the LQCD calculation at the same $L$, since
the information of the nucleon resonances embedded in the data have been
coded in the constructed Hamiltonian.
No need to perform  expensive LQCD calculations covering  a wide range of $L$  
for investigating the nucleon resonances, such as the Roper $N^*(1440)$ resonance,
 which decay into multi-channel states.

\begin{figure}[htbp] \vspace{-0.cm}
\begin{center}
\includegraphics[width=0.45\columnwidth]{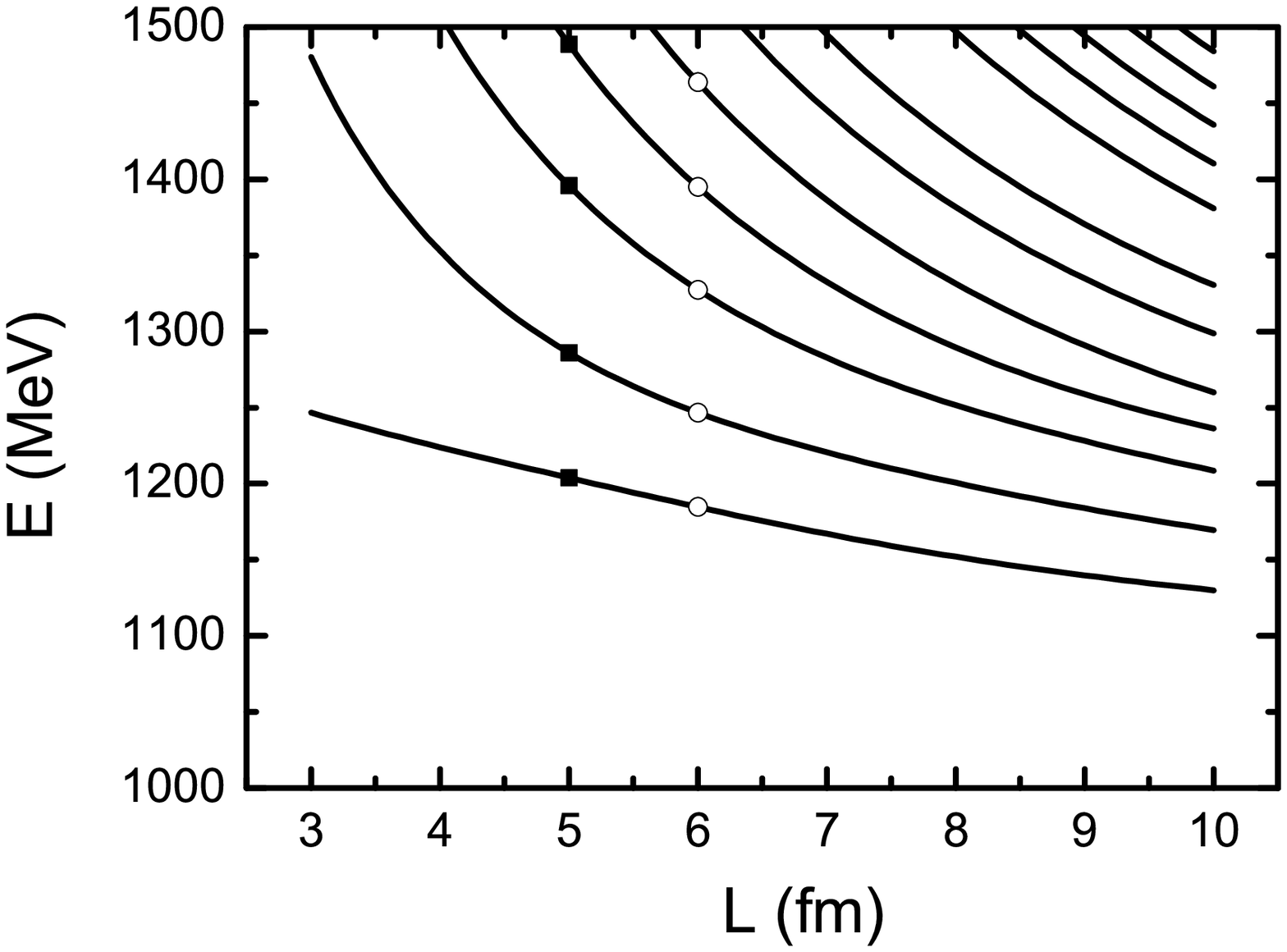}
\includegraphics[width=0.45\columnwidth]{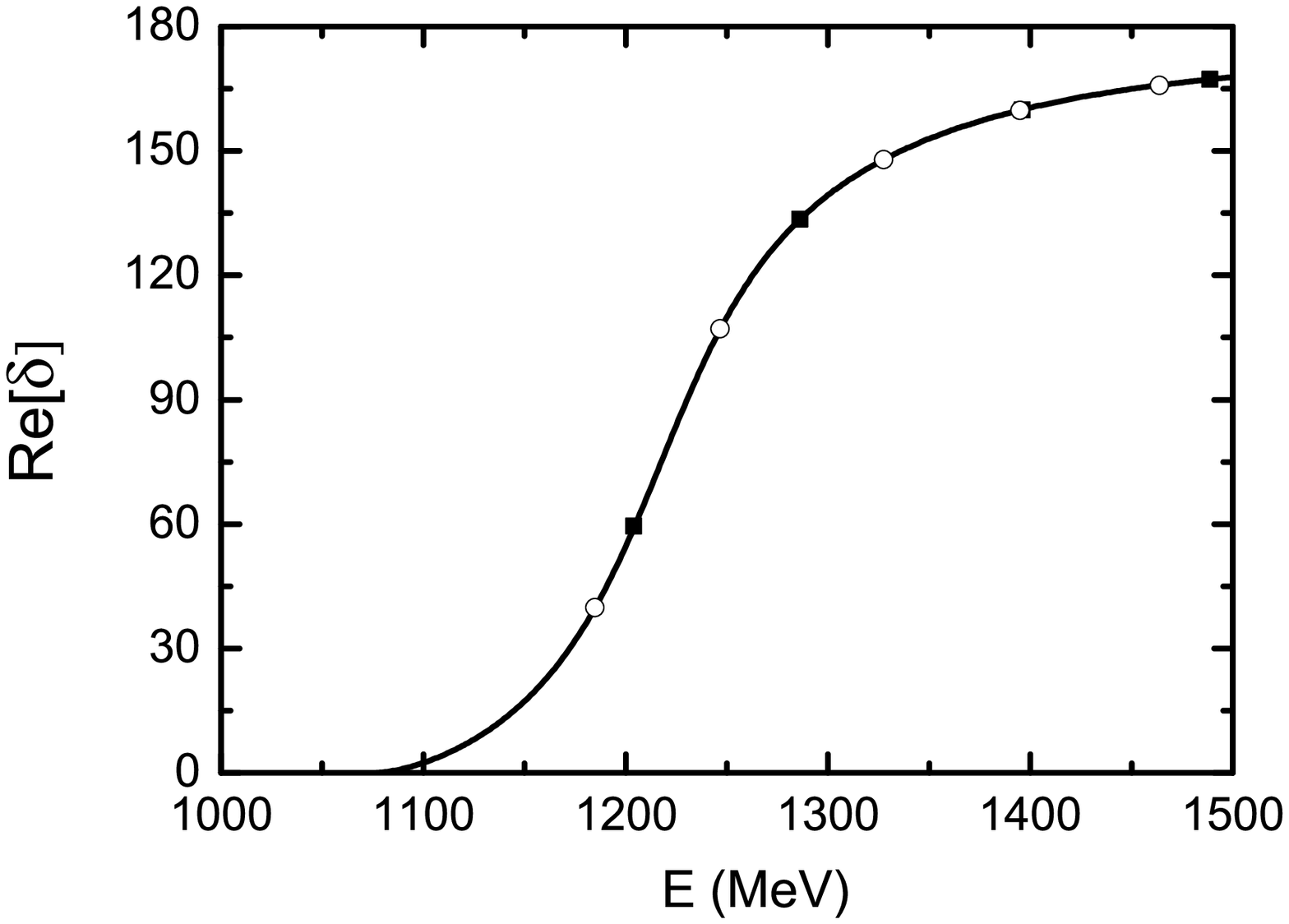}
\caption{ The spectrum (left) and phase shifts (right) in $P_{33}$ state
of $\pi N$ system calculated 
from using the SL model\cite{sl}.}
\label{fg:sl}
\end{center}
\end{figure}

The finite-volume Hamiltonian method was established in Ref.\cite{ad-1} using
a simple one-channel ($n_c=1$) Hamiltonian consisting of a $\Delta\rightarrow \pi N$ vertex 
and a separable $\pi N\rightarrow \pi N$ separable potential.
Here we confirm the method by using the meson-exchange model (SL model)
of Ref.\cite{sl}. The results are shown in Fig.\ref{fg:sl}.
The predicted energy levels as function of the volume size $L$ are shown in
the left-hand side, and  the solid curve in the 
right-hand side is drawn from using
the phase shifts calculated from the SL model in infinite volume. For each energy $E$ 
at a given $L$ in the left-hand side,  the L\"uscher formula can be used to 
get the phase shift $\delta(E)$ by
\begin{eqnarray}
\delta(E)&=&-tan^{-1}(-\frac{q\pi^{3/2}}{Z_{00}(1;q^2)})+ n\pi\,,
\label{eq:luch1}
\end{eqnarray}
where $q=\frac{kL}{2\pi}$ is evaluated by the momentum $k$ defined by
 $E=E_N(k)+E_\pi(k)$, and 
$Z_{00}(1;q^2)$ is the generalized Zeta function.
The phases calculated from each energy at $L=5,6$ in the left-hand side of Fig.\ref{fg:sl} are
the points in the right-hand side, which agree with the solid curve.
Thus the finite-volume Hamiltonian method is equivalent to the use of
 L\"uscher formula to relate the spectrum of finite volume to the scattering 
amplitudes which are obtained from fitting the experimental data through 
the SL model Hamiltonian.

The L\"uscher's
formula for the cases with two open channels, such as that derived in Ref.\cite{luch-2}, can be written as
\begin{eqnarray}
\cos\left[\phi(q_{1}(L))+\phi(q_2(L))-\delta_{1}-\delta_2\right]
%\nonumber \\
\quad-\eta\cos\left[\phi(q_1(L))-\phi(q_{2}(L))-\delta_{1}+\delta_{2}\right]=0\,,
\label{eq:luch2a}
\end{eqnarray}
where $\phi(q_{i}(L)=tan^{-1}(-\frac{q_i(L)\pi^{3/2}}{Z_{00}(1;q^2_i(L))})$ with
$q_{i}(L)=\frac{k_i(E)L}{2\pi}$, $\delta_1(E)$ ($\delta_2(E))$ 
is the phase shifte
for channel 1 (2), and $\eta$ is the inelasticity.
In the rest frame, this means that we need to perform calculations for
three different values of $L$ if Eq.(\ref{eq:luch2a}) is used to test LQCD results
against the  data of two phase shifts and inelasicity.
On the other hand, the spectrum from finite-volume Hamiltonian method at only one $L$
is sufficient to test LQCD calculations at the same $L$ since  
the resonances embedded in the data have been coded in the Hamiltonian.
Alternatively, one can use the LQCD spectrum to  construct a K-matrix model, such as that
done in Ref.\cite{jlab-lqcd},
and then look for the resonance poles. From the results discussed in the previous section,
it is clear that such an approach will be reliable only when 
 the predicted phase shifts and inelasticity are of high accuracy and cover
 a sufficiently wide  range of  energies such that  the parameters of the
K-matrix are well conatrained. Thus the
LQCD calculations for a wide range of $L$ are required.   
Here we see the advantage of using the finite volume Hamiltonian method over the use
of L\"uscher formula to test LQCD calculations.

The ANL-Osaka model Hamiltonian can be schematically written
as $H=H_0+H_I$, where $H_0$ is the free Hamiltonian and the interaction Hamiltonian
can be written as
\begin{eqnarray}
H_I= \sum_{i}\sum_{\alpha}g_{N^*_i, \alpha}
+\sum_{M=\rho,\sigma}f_{M,\pi\pi} + \sum_{\alpha,\beta} v_{\alpha,\beta}\,,
\label{eq:anl-osaka-h}
\end{eqnarray}
where $\alpha,\beta = \pi N, \eta N, K\Lambda, K\Sigma, \pi\pi N(\pi \Delta, \sigma N, \rho N)$,
$g_{N^*_i, \alpha}$ defines the decay of the $i$-th bare $N^*$ state into channel $\alpha$,
$v_{\alpha,\beta}$ denotes the meson-exchange potential between
channels $\alpha$ and $\beta$, and $f_{M,\pi\pi}$ describes the decay of meson $M$ into
$\pi\pi$. The interactions are determined by fitting very extensive data
of $\pi N,\gamma N\rightarrow \pi N, \eta N, K\Lambda, K\Sigma, \pi\pi N$ processes
and the nucleon resonances have been extracted.
Thus,  a  LQCD calculation can be related to the
nucleon resonances embedded in the $\pi N$ and $\gamma N$ reactions
if its predicted   spectrum  agree with the spectrum calculated
from the ANL-Osaka model Hamiltonian by solving Eq.(\ref{eq:det-1})  in finite volume.

In the presence of the transitions to $\pi\pi N$ states due
to the $\Delta \rightarrow \pi N$ and $\rho,\sigma \rightarrow \pi\pi$ 
vertex interactions in Eq.(\ref{eq:anl-osaka-h}), 
it is rather complex to apply the finite-volume
Hamiltonian method to the ANL-Osaka Hamiltonian.  
To compare our approach with the approach using the 
L\"uscher formula,
it is sufficient to consider a three-channel Hamiltonian which
is deduced from Eq.(\ref{eq:anl-osaka-h})
 by keeping only $\pi N$, $\pi\Delta$, and $\sigma N$ channels
and neglecting the $f_{\rho,\pi\pi}$ and $f_{\sigma,\pi\pi}$ vertex interactions.
We determine this three-channel model Hamiltonian by fitting the
$\pi N$ scattering amplitudes up to only $1.6$ GeV. Except the $S_{11}$ partial
wave, which is known to have large coupling with the  $\eta N$ channel and therefore
cannot be fitted well here, 
the fits are comparable to those of the  ANL-Osaka results.
The extracted resonance poles are with masses $M_R =1353.5\,\, - i\,38.3$ MeV for $P_{11}$ and
$M_R=1211.9\,\,-i\, 52.8$ MeV for $P_{33}$. The value for $P_{33}$ is close to the
$M_R=1216.4\,\,-i\, 50.0$ MeV of SL model\cite{sl}. This is consistent with
what we have shown in the previous section, since the constructed 3-channel model
and the SL model give almost
the same fits to $P_{33}$ amplitude data below 1.3 GeV.

\begin{figure}[htbp] \vspace{-0.cm}
\begin{center}
\includegraphics[width=0.7\columnwidth]{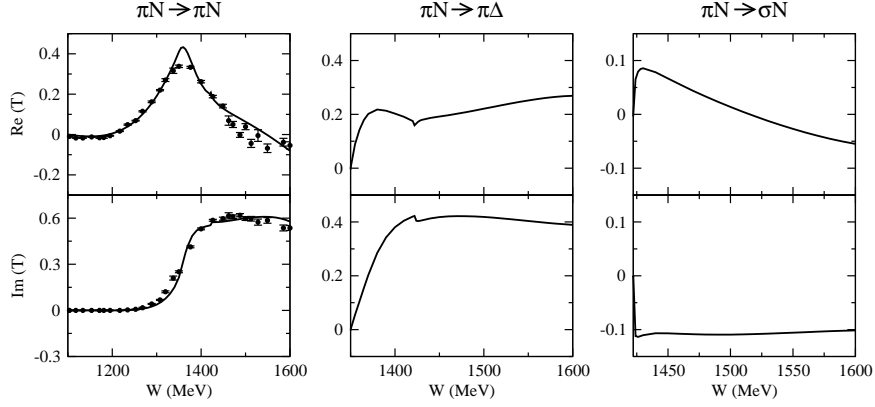}
\caption{ The $\pi N \rightarrow \pi N$ (left),
$\pi N \rightarrow \pi \delta$ (middle), and $\pi N\rightarrow \sigma N$ (right)
amplitudes for the $P_{11}$ state calculated from the 3-channel model described in
the text.}
\label{fg:amp-p11}
\end{center}
\end{figure}

For our discussion here, we  show in Fig.\ref{fg:amp-p11} the resulting amplitudes for
the $ \pi N \rightarrow \pi N, \pi \Delta, \sigma N$ transitions in the $P_{11}$ partial wave.
These three amplitudes contain the information determined by the data 
of $\pi N \rightarrow \pi N, \pi \pi N $ reactions. Thus a LQCD calculation 
aimed at investigating the Roper $N^*(1440)$ should at least be consistent with these
three amplitudes, since it is known that the decay width for 
$N^*(1440) \rightarrow \pi \Delta, \sigma N\rightarrow \pi \pi N$ is very large.
This can only be achieved by using  either the multi-channel L\"uscher formula\cite{luch-3}
or the finite-volume Hamiltonian method. We now turn to comparing these two
different approaches within the
three-channel model described above.

\begin{figure}[htbp] \vspace{-0.cm}
\begin{center}
\includegraphics[width=0.7\columnwidth]{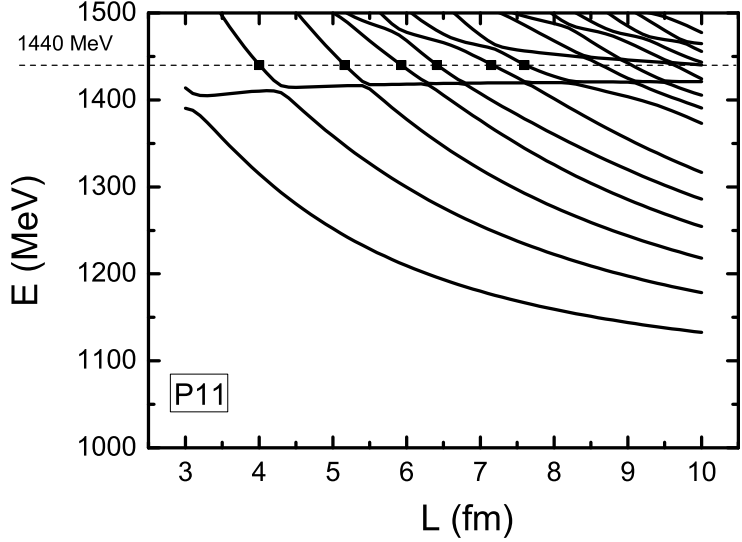}

\caption{The spectrum for the $P_{11}$ state 
calculated from from the 3-channel model described in
the text.}
\label{fg:spect-p11}
\end{center}
\end{figure}

For the three-channel model with $n_c=3$, the matrix for the free
Hamiltonian in Eq.(\ref{eq:det-1}) takes the following form
\begin{eqnarray}
[H_0]_{3N+1}&=&\left( \begin{array}{cccccccc}
m_0                     & 0                            & 0
                        & 0                            & 0  & \cdots \\
0                       & E_\pi(k_0)+E_N(k_0)      & 0
                        & 0                            & 0  & \cdots \\
0                       & 0                            & E_\pi(k_0)+E_\Delta(k_0)
                        & 0                            & 0  & \cdots \\
0                       & 0                            & 0
                        & E_\sigma(k_0)+E_N(k_0)      & 0  & \cdots \\
0                       & 0                            & 0
                        & 0                            & E_\pi(k_1)+E_N(k_1) & \cdots \\
\vdots                  & \vdots                       & \vdots
                        & \vdots                       & \vdots                & \ddots
\end{array} \right), \nonumber
\end{eqnarray}
where $E_a(k)=\sqrt{m_a^2+k^2}$ is the energy of particle $a$ with a mass $m_a$,
The $(3N+1)\times(3N+1)$ matrix for the interaction Hamiltonian is
\begin{eqnarray}
[H_I]_{3N+1}&=&\left( \begin{array}{cccccccc}
0                       & g^{fin}_{\pi N}(k_0)               & g^{fin}_{\pi\Delta}(k_0)
                        & g^{fin}_{\sigma N}(k_0)               & g^{fin}_{\pi N}(k_1)              & \cdots \\
g^{fin}_{\pi N}(k_0)   & v^{fin}_{\pi N,\pi N}(k_0, k_0)   & v^{fin}_{\pi N,\pi \Delta}(k_0, k_0)
                        & v^{fin}_{\pi N,\sigma N}(k_0, k_0)   & v^{fin}_{\pi N,\pi N}(k_0, k_1)   & \cdots \\
g^{fin}_{\pi\Delta}(k_0) & v^{fin}_{\pi\Delta,\pi N}(k_0, k_0) & v^{fin}_{\pi\Delta,\pi\Delta}(k_0, k_0)
                        & v^{fin}_{\pi\Delta,\sigma N}(k_0, k_0) & v^{fin}_{\pi\Delta,\pi N}(k_0, k_1)   & \cdots \\
g^{fin}_{\sigma N}(k_0)   & v^{fin}_{\sigma N,\pi\pi}(k_0, k_0)   & v^{fin}_{\sigma N,\pi\Delta}(k_0, k_0)
                        & v^{fin}_{\sigma N,\sigma N}(k_0, k_0)   & v^{fin}_{\sigma N,\pi N}(k_0, k_1)   & \cdots \\
g^{fin}_{\pi N}(k_1) & v^{fin}_{\pi N,\pi N}(k_1, k_0) & v^{fin}_{\pi N,\pi\Delta}(k_1, k_0)
                        & v^{fin}_{\pi N,\sigma N}(k_1, k_0) & v^{fin}_{\pi N,\pi N}(k_1, k_1)   & \cdots \\
\vdots                  & \vdots                       & \vdots
                        & \vdots                       & \vdots                  & \ddots
\end{array} \right), \label{eq:detf}
\end{eqnarray}
with
\begin{eqnarray}
g^{fin}_{\alpha}(k_n)&=&\sqrt{\frac{C_3(n)}{4\pi}}\left(\frac{2\pi}{L}\right)^{3/2}
g_{N^*_1,\alpha}(k_n),\\
v^{fin}_{\alpha,\beta}(k_{n_i},k_{n_j})&=&\sqrt{\frac{C_3(n_i)}{4\pi}}
\sqrt{\frac{C_3(n_j)}{4\pi}}\left(\frac{2\pi}{L}\right)^3 v_{\alpha,\beta}(k_{n_i},k_{n_j}),
\label{eq:vfin-1}
\end{eqnarray}
where $C_3(n)$ is the number of degenerate states with the same magnitude $k_n=|\vec{k}_n|$.
By solving Eq.~(\ref{eq:det-1}), we then obtain the spectra for 
$P_{11}$ shown in
Fig.~\ref{fg:spect-p11}.

By choosing the normalization to relate the T-matrix elements to S-matrix elements by
 $S_{\alpha,\beta}(E)=\delta_{\alpha,\beta} - 2i T_{\alpha,\beta}(E)$,
the L\"uscher formula given in  Ref.\cite{luch-3} for 
the constructed 3-channel model can be written
explicitly as :
%\begin{eqnarray}
%\det[M(E,L)]=0\,,
%\label{eq:detf}
%\end{eqnarray}
%where
\begin{eqnarray}
\det\left( \begin{array}{ccc}
T_{\pi N,\pi N}(E)+C_{\pi N,\pi N}(L,E)            & T_{\pi N,\pi\Delta}(E)
                            & T_{\pi N,\sigma N}(E) \\
T_{\pi \Delta, \pi N}(E)                       & T_{\pi \Delta,\pi\Delta}(E)+C_{\pi\Delta,\pi\Delta}(L,E)
      & T_{\pi\Delta,\sigma N}(E) \\
T_{\sigma N, \pi N}(E)                       & T_{\sigma N,\pi \Delta}(E)
          & T_{\sigma N,\sigma N}(E)+C_{\sigma N,\sigma N}(L,E)
\end{array} \right)=0\,, \nonumber
\end{eqnarray}
with
\begin{eqnarray}
 C_{\alpha,\alpha}(L,E)&=&iq_\alpha(L)/(q_\alpha(L)-4\sqrt{\pi}Z_{00}(1;q_\alpha(L)),
\label{eq:detf-1}
\end{eqnarray}
and
$q_\alpha(L)=k_\alpha L/(2\pi)$ defined by the on-shell momentum $k_\alpha$ of the channel $\alpha$.

Because of symmetries and the unitarity conditions, only six  of the total 12
real numbers that are needed to specify all of  the six complex
$T_{\alpha,\beta}(E)$ matrix elements are independent. Thus we need to get six relations from
Eqs.(\ref{eq:detf})-(\ref{eq:detf-1}) at each $E$ to relate the spectrum to  the
scattering amplitudes shown in
Fig.\ref{fg:amp-p11} for $P_{11}$.
 In the rest frame, this means that we need to perform LQCD calculations
at 6 different $L$. For $E=1440$ MeV, these are 
the 6 interaction points (solid squares)
 between the dashed line and the solid curves in  Fig\ref{fg:spect-p11}.
Clearly, this will be a very difficult, if not impossible, LQCD calculation.
On the other hand, the information on the Roper $N^*(1440)$ resonance has been coded in the
3-channel Hamiltonian by fitting  the empirical $\pi N$ scattering amplitudes\cite{said}
as shown in Fig.\ref{fg:amp-p11}. Therefore the spectrum from finite-volume Hamiltonian method
at $any$ given L
is sufficient to test LQCD calculation. Here we see the great advantage  of
the finite-volume Hamiltonian method over the approach using the L\"uscher formula to test LQCD
calculations aimed at investigating the nucleon resonances.

\section{Summary}
By using several exactly soluble $\pi\pi$ scattering models. 
we have shown  that
in reality the determinations of PWA and resonance
extractions cannot be performed model independently.
It is desirable to extract the  nucleon resonances  within a reaction model
that is constrained by the
well-established physics, such as the meson-exchange mechanisms included
in the ANL-Osaka and J\"uelich analyses. 

Within a three-channel
model with $\pi N$, $\pi\Delta$ and $\sigma N$ channels, we show the advantage
of the finite-volume Hamiltonian method over the approach using the L\"uscher formula
to test Lattice QCD calculations aimed at investigating 
 the properties of excited nucleon states.
To apply the finite-volume Hamiltonian method to predict spectra using
 the ANL-Osaka dynamical multi-channel Hamiltonian, we need to develop approaches to handle
$\pi\pi N$ channels.
Since the information on about 25 nucleon resonances with mass up to 2 GeV extracted from the
very  extensive data of $\pi N, \gamma N \rightarrow \pi N, \eta N, K\Sigma, K\Sigma, \pi\pi N$
reactions have been coded
in the ANL-Osaka model Hamiltonian, the predicted spectra can readily be used to test the
LQCD calculations aimed at investigating the structure of the excited nucleons.  Furthermore,
the ANL-Osaka analysis will be applied to include new data from 12 GeV upgrade experiments and
thus it will provide more accurate information for testing LQCD calculations in the near
future.

\begin{acknowledgements}
\vspace{1cm}
This work was supported by the U.S. Department of Energy, Office of Science, Office of Nuclear Physics, Contract No. DE-AC02-06CH11357.
This research used resources of the National Energy Research Scientific Computing Center,
which is supported by the Office of Science of the U.S. Department of Energy
under Contract No. DE-AC02-05CH11231, and resources provided on Blues and/or Fusion,
high-performance computing cluster operated by the Laboratory Computing Resource Center
at Argonne National Laboratory.
\end{acknowledgements}

%\clearpage


\begin{thebibliography}{3}
\bibitem{anl-osaka}
H. Kamano, S.X. Nakamura, T.-S. H. Lee, and T. Sato, Phys. Rev. C {\bf 88},035209 (2013).
\bibitem{bg}
A. V. Anisovich, R. Beck, E. Klempt, V. A. Nikonov, A. V. Sarantsev, and U. Thoma, Eur. Phys. J. A 
48, 15 (2012).
\bibitem{juelich}
 D. Ro ̈nchen, M. D ̈oring, F. Huang, H. Haberzettl, J. Haidenbauer, C. Hanhart, S. Krewald,
U.-G. Meissner, and K. Nakayama, Eur. Phys. J. A 49, 44 (2013).
\bibitem{ad-1}
 J.~M.~M.~Hall, A.~C.-P.~Hsu, D.~B.~Leinweber, A.~W.~Thomas and R.~D.~Young,
 Phys.\ Rev.\ D {\bf 87}, 094510 (2013).
\bibitem{ad-2}
Jia-Jun Wu, T.-S.H. Lee, A.W. Thomas, R.D. Young, Phys.Rev. C {\bf 90}, 055206 (2014)
\bibitem{dalitz}

R. H. Dalitz and R. G. Moorhouse, Proc. Roy. Soc. Lond. A318, 279 (1970); A. J. F. Siegert,
Phys. Rev. 56, 750 (1939).
morehouse
\bibitem{bohm}
A. Bohm, Quantum mechanics: foundations and applications (Springer-Verlag, New York,
1993).

\bibitem{tabakin}
W.-T. Chiang and F. Tabakin, Phys. Rev. C 55 (1997) 2054.
\bibitem{shkl}
A. M. Sandorfi, S. Hoblit, H. Kamano, and T.-S. H. Lee, J. Phys. G 38, 053001 (2011).
\bibitem{msl}
 A.~Matsuyama, T.~Sato and T.~-S.~H.~Lee, Phys.\ Rept.\  {\bf 439}, 193 (2007)
\bibitem{mach}
R. Machleidt, Chapter 2, Vol.19, Advances in Nuclear Physics, eds. J.W. Negele and E. Vogt, Plenum Press (1989).
\bibitem{sl}
T. Sato and T.-S. H. Lee, Phys. Rev. C {\bf 54}, 2660  (1996);
Phys. Rev. C {\bf 63}, 055201 (2001).

\bibitem{luch-1}
  M.~L\"uscher, Nucl.\ Phys.\ B {\bf 354}, 531 (1991).

\bibitem{luch-2}
  S.~He, X.~Feng and C.~Liu, JHEP {\bf 0507}, 011 (2005)

\bibitem{luch-3}
  M.~T.~Hansen and S.~R.~Sharpe, Phys.\ Rev.\ D {\bf 86}, 016007 (2012)
\bibitem{jlab-lqcd}
David J. Wilson, Raul A. Briceno, Jozef J. Dudek, Robert G. Edwards, Christopher E. Thomas, Phys. Rev. D {\bf 92}, 094502 (2015)  
\bibitem{said}
CNS Data Analysis Center, George Washington University,
http://gwdac.phys.gwu.edu

\end{thebibliography}
\end{document}